\documentclass[reprint,showkeys,aip,apl,numerical,twocolumn]{revtex4-1}
\usepackage{amsmath}
\usepackage{color}
\usepackage{hyperref}
\usepackage{graphicx}
\usepackage{setspace}
\usepackage{comment} 

\newcommand{\VEC}[1]{{\boldsymbol{ #1}}}

\newcommand{\GP}{{Gr\"uneisen parameter}}
\newcommand{\GPs}{{Gr\"uneisen parameters}}
\newcommand{\Gru}{{Gr\"uneisen}}
\newcommand{\Fig}{{Fig.}}
\newcommand{\sbte}{{Sb$_2$Te$_3$}}
\newcommand{\bise}{{Bi$_2$Se$_3$}}
\newcommand{\invcm}{{cm$^{-1}$}}

\begin{document}
\title{
Anharmonic phonon effects on linear thermal expansion of trigonal bismuth selenide and antimony telluride crystals
}
\author{Chee Kwan Gan}
\email{ganck@ihpc.a-star.edu.sg}
\affiliation{Institute of High Performance Computing, 1 Fusionopolis Way, \#16-16 Connexis, Singapore 138632}
\author{Ching Hua Lee}
\affiliation{Institute of High Performance Computing, 1 Fusionopolis Way, \#16-16 Connexis, Singapore 138632}
\date{15 March 2018 (b: CMS version)} 

\begin{abstract}
We adopted and extended an efficient \Gru{} formalism
to study the phonon anharmonicity and linear thermal expansion
coefficients (TECs) of trigonal bismuth selenide (\bise{}) and antimony
telluride (\sbte{}). Anharmonicity of the systems is studied via extensive
calculation of \GPs{} that exploit symmetry-preserving deformations.  Consistent with experimental findings, a large anisotropy between the TECs in
the $a$ and $c$ directions is found.
The larger anharmonicity inherent in \sbte{}, as compared to \bise{} is
offset by the volumetric effect, resulting in comparable temperature dependence of their linear TECs.
The Debye temperatures deduced from our first-principles data also agree very well with the existing tabulated
values.
The highly efficient methodology developed in this work, applied for the first time to study the linear TECs of two trigonal 
thermoelectric systems, opens up exciting opportunities to address the 
anharmonic effects in other thermoelectrics and other low-symmetry materials.
\end{abstract}

\maketitle

\section{Introduction}
Bismuth selenide (\bise) and antimony telluride (\sbte) belong to a
large family of metal dichalcogenides that hosts excellent thermoelectric
materials\cite{Snyder08v7} and topological
insulators\cite{Hasan10v82,Zhang09v5,Zhang10v12}.  As paradigmatic 
examples of materials that simultaneously host enigmatic 3D $Z_2$ topological
states, these two materials have been extensively
studied experimentally\cite{Chen11v99,Dutta12v100,Tian17v95,Das17v96} and theoretically\cite{Sosso09v21,Bessas12v86}
due to their
technological importance and fundamental interest. 
The linear and
volumetric thermal expansion coefficients (TECs) of  \bise{} and \sbte{}
have been determined experimentally\cite{Chen11v99} where a high
anisotropy is found between linear TECs in the $a$ and $c$ directions
for these two systems. 

For engineering applications of these materials, good device performance hinges on a solid understanding of thermal expansion behavior because phonon dynamics is intimately affected by temperature-induced crystal deformations. As found in \cite{Lin11v83,Kim12v100}, knowledge of the linear thermal expansion and phonon anharmonicity can be captured through phonon frequency lineshifts through the \GPs{}.  Such calculations of the thermal expansion properties are commonly performed using a quasi-harmonic
approximation (QHA), which involves many phonon calculations on many possible combinations of lattice parameters. But due to its complexity, the QHA is efficient only when dealing with highly symmetric systems such as cubic lattice structures. However,
many technological important crystals are not cubic, and other more efficient approaches are necessary. In this paper, we adopted and extended an efficient \Gru{}
approach by\cite{Schelling03v68,Mounet05v71,Gan15v92,Gan16v94} to
study \bise{} and \sbte{} with a minimal set of
relatively expensive (compared to standard density-functional total-energy
calculations) phonon calculations. Through it, we were able to perform a systematic investigation on the anharmonicity of these two materials with relatively low symmetry, and make consistent comparisons between some of their important thermal properties such as linear TECs.

\section{Methodology}
The trigonal \bise{} and \sbte{} belong to the symmorphic space group
$R{\overline 3}m$ (No. 166).  There are three inequivalent atoms: an
Sb atom occupies $6c(0,0,\mu)$ site, a Te atom occupies $3a(0,0,0)$
site, and a second Te atom occupies $6c(0,0,\nu)$ site. This gives a
total of $15$ atoms in the conventional hexagonal unit cell.  However,
in order to reduce the amount of computing time, we use a primitive
rhombohedral cell of five atoms that is three times smaller than the conventional
hexagonal cell.  The rhombohedral cell length $a_r$
and angle $\alpha_r$ can be deduced from the hexagonal lattice parameters
$a_h$ and $c_h$, and vice versa.  The relations are: $a_h = 2 a_r \sin
(\alpha_r/2)$, $c_h = a_r \sqrt{ 3 + 6 \cos\alpha_r}$.  On the other
hand, $a_r = (a_h/3)\sqrt{\eta^2+ 3}$, $\cos\alpha_r = ( 2\eta^2 -
3   )/(  2 \eta^2 + 6  )$ where $\eta = c_h/a_h$.

We perform density-functional theory (DFT) calculations within the
local density approximation as implemented in the plane-wave basis
suite QUANTUM ESPRESSO\cite{Giannozzi09v21} (QE), with wavefunction
and density cutoffs of $60$ and $480$~Rydberg, respectively. A
$10\times 10 \times 10$  Monkhorst-Pack mesh is used for the
$k$-point sampling. The pseudopotentials for Bi, Se, Sb, and
Te are generated using the pslibrary.1.0.0 that is based on the
Rappe-Rabe-Kaxiras-Joannopoulos\cite{Rappe90v41} scheme.  We relax the
structures fully before carrying out the phonon
calculations.  For \bise{}, we obtain $(a,c) = (4.110, 27.900)$~\AA. This
is in good agreement with the experimental\cite{Nakajima63v24} result
of $(4.143,28.636)$~\AA{}.
For \sbte{}, we obtain $(a,c) = (4.244, 29.399)$~\AA{}, which is in
good agreement with the experimental\cite{Chen11v99} result of $(4.242,
30.191)$~\AA.

According to the \Gru{}
approach\cite{Pavone93v48,Schelling03v68,Mounet05v71,Gan15v92,Gan16v94,Lee17v96},
the linear TECs in the $a$ and $c$ directions, denoted as $\alpha_a(T)$
and $\alpha_c(T)$, respectively, are given by \begin{equation}
\begin{pmatrix} \alpha_{a}\\ \alpha_{c} \end{pmatrix} = \frac{1}{\Omega
D} \begin{pmatrix} C_{33} & -C_{13} \\ -2C_{13} & [C_{11} +C_{12}] \\
\end{pmatrix} \begin{pmatrix} I_{1}\\ I_{3} \end{pmatrix} \label{eq:TECs}
\end{equation} where $D = (C_{11} + C_{12}) C_{33} - 2C_{13}^2$. For
clarity, the explicit dependence of $\alpha$'s and $I_i$'s on
temperature $T$ is suppressed in Eq.~\ref{eq:TECs}.  We will discuss
more about $I_i$ later. The $C_{ij}$ are
the elastic constants. The linear TECs  are inversely proportional to
the volume $\Omega$ of primitive cell at equilibrium.  We note that
\bise{} has a smaller $\Omega$ than \sbte{} (i.e., $ 136.05$~\AA$^3$ vs
$152.87$~\AA$^3$).  From a series of symmetry-preserving deformations with
strain parameters ranging from $-0.01$ to $0.01$, the elastic constants
are deduced from parabolic fits to the energy-strain\cite{DalCorso16v28}
curves.  For \bise{}, $C_{11} + C_{12} = 121.74$, $C_{13}  = 30.18$,  and
$C_{33} = 54.45$~GPa.  For \sbte{}, $C_{11} + C_{12} = 110.73$, $C_{13} =
32.16$, and $C_{33} = 60.97$~GPa.  We note that the expression for TECs
in Eq.~\ref{eq:TECs} is identical to the hexagonal case\cite{Gan16v94}
since a trigonal cell can be perfectly embedded in a hexagonal cell.

\begin{widetext}

\begin{figure}[htbp]
\centering\includegraphics[width=17.2cm,clip]{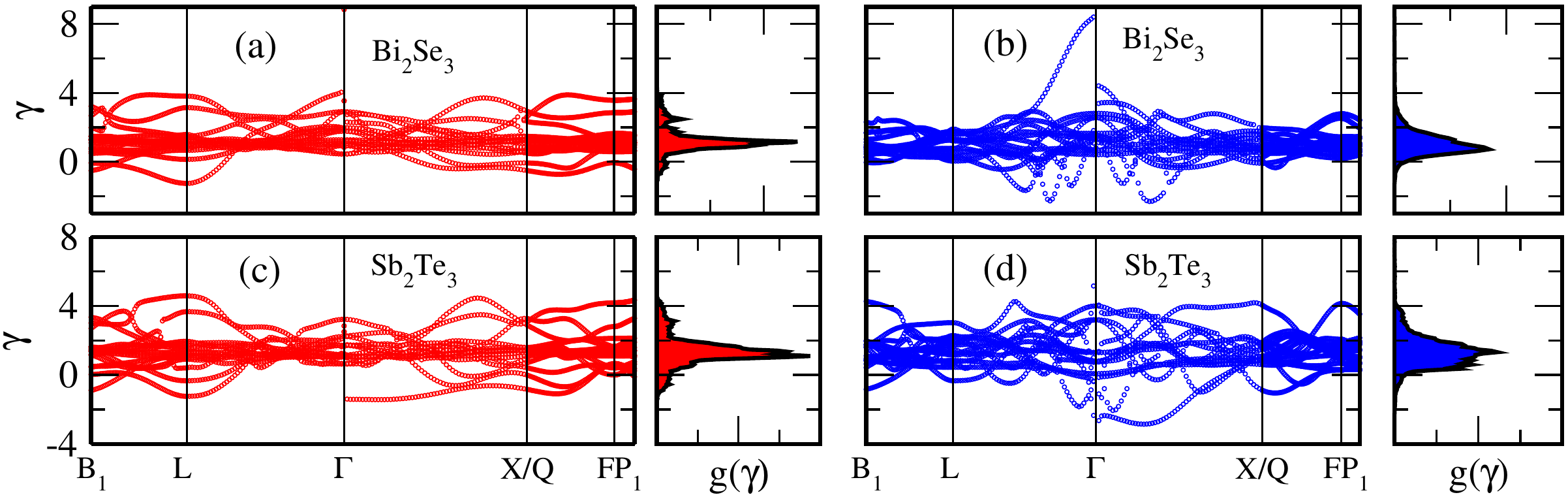}
\caption{
The \GPs{} for \bise{} due to (a) an $xy$ biaxial strain and (b) a $z$
uniaxial strain. The corresponding results for \sbte{} are shown in (c)
and (d), respectively. The label and coordinates of the $k$ points are
taken from Ref.~[\onlinecite{Setyawan10v49}]. The densities of \GPs{},
$g(\gamma)$, shown on the right side of each figure are obtained with
a sampling of $30\times 30\times 30$ $k$ points.
}
\label{fig:GP}
\end{figure}

\end{widetext}

Central to our \Gru{} formalism is 
the temperature dependent heat capacity weighted by the \GP{},
\begin{equation}
I_i(T) = 
\frac{\Omega}{(2\pi)^{3}}
\sum_{\lambda} \int_{\rm BZ} d\VEC{k}\ \gamma_{ i, \lambda\VEC{k} } c( \nu_{\lambda \VEC{k}},T)
\label{eq:dsumI}
\end{equation}
where the integral is over the first Brillouin zone (BZ).  Here $
\gamma_{ i, \lambda\VEC{k} } = - n^{-1} \nu_{\lambda \VEC{k}}^{-1}
\partial  \nu_{\lambda \VEC{k}} / \partial \epsilon_i $ are the
mode-dependent and deformation-dependent \GPs, which measures the rate
of change of the phonon frequency $\nu_{\lambda \VEC{k}}$ (of mode
index $\lambda$ and wavevector $\VEC{k}$) with respect to the strain
parameter $\epsilon_i$. $n$ equals to 1 (2) for a uniaxial (biaxial)
strain.  The specific heat contributed by a phonon mode of frequency $
\nu$ is $c(\nu,T) = k_B(r/\sinh r)^2$, $r=h \nu/2k_B T$. $k_B $ and $h$
are the Boltzmann and Planck constants, respectively.  
To keep track of the origin of anharmonicity more precisely, we further define the 
the density of phonon states weighted by \GP{}, $\Gamma_i(\nu) $, given by
\begin{equation}
\Gamma_{i}(\nu) =
\frac{\Omega}{(2\pi)^{3}} \sum_{\lambda} \int_{\rm BZ} d\VEC{k}\ \delta(\nu
- \nu_{\lambda\VEC{k} }) \gamma_{ i, \lambda\VEC{k} } 
\end{equation}
such that 
$I_i(T) = \int_{\nu_{\rm min}}^{\nu_{\rm max}} d\nu\ \Gamma_i(\nu) c(\nu,T)$. $\nu_{min}$ ($\nu_{max}$) is the minimum (maximum) frequency
in the phonon spectrum.  
The functions $\Gamma_i(\nu)$ provide a deeper understanding
about $I_i(T)$ since it isolates the anharmonicity-dependent
contributions from the harmonic specific heat capacity $c(\nu,T)$, which has a well-known universal form\cite{Gan15v92}.  
Finally we note that $I_i(T) $ is
related to the macroscopic\cite{AshcroftMermin-book} \GPs{}, $\gamma_{m,i}(T) $ by
the relation $\gamma_{m,i}(T) =  I_i(T)/C_v(T) $ where $C_v(T) =
\frac{\Omega}{(2\pi)^{3}} \sum_{\lambda} \int_{\rm BZ} d\VEC{k}\ c(
\nu_{\lambda \VEC{k}},T)  $ is the specific heat at constant volume.
Therefore $\gamma_{m,i}(T)$ can be interpreted as an average over \GPs{}
weighted by the mode dependent heat capacity. Its physical meaning is
clearest in the large-$T$ limit, where $\gamma_{m,i}$ reduces to a simple
arithmetic average of all \GPs{} in the BZ since the heat capacities
for each mode approaches unity (in units of $k_B$) in this limit.

To calculate the \GPs{} resulted from a deformation of the
crystal\cite{Lee17v96} due to an $xy$ biaxial strain, a strain-parameter set of
$(\epsilon_1,\epsilon_1,0,0,0,0)$ (in Voigt's notation) is used,
where the rhombohedral cell has a new lattice parameters $a_r' = a_r
\sqrt{[\eta^2 + 3(1+\epsilon_1)^2]/(\eta^2 + 3)}$ and $\cos \alpha_r' =
[ 2\eta^2 - 3(1+\epsilon_1)^2 ]/[ 2\eta^2 + 6(1+\epsilon_1)^2   ]$.
For  a $z$ uniaxial strain, we use the strain-parameter set of
$(0,0,\epsilon_3,0,0,0)$, where the rhombohedral cell has $a_r' = a_r
\sqrt{[\eta^2(1+\epsilon_3)^2 + 3]/(\eta^2 + 3)}$ and $\cos \alpha_r'
= [ 2\eta^2(1+\epsilon_3)^2 - 3  ]/[ 2\eta^2(1+\epsilon_3)^2 + 6 ]$.
Importantly, these two deformations preserve the space group of the crystal so that
we can use the QE symmetry switch of IBRAV=5.  We use small strains of
$e_1 = \pm 0.25~\%$ and $e_3 = \pm 0.5~\%$ for the calculation \GPs{} using finite-differences.
For phonon calculations under the QE implementation, we use a $q$ mesh
of $5 \times 5 \times 5$, which is equivalent to a $5\times 5 \times 5$
supercell\cite{Liu14v16} for the determination of interatomic force
constants.

\section{Results}

The \GPs{} along the representative high-symmetry directions for
\bise{} and \sbte{} due to an $xy$ biaxial strain are shown in
\Fig~\ref{fig:GP}(a) and (c), respectively.  Similarly, the results
due to a $z$ uniaxial strain are shown in \Fig~\ref{fig:GP}(b) and (d),
respectively. The densities of \GPs{} (displayed on the right side of each
subfigure) show that most \GPs{} range between 0 to 4, with a dominant
peak centered around 1. There is a small population of negative \GPs{},
which may lead to negative linear TECs\cite{Mounet05v71}.

\begin{figure}[htbp]
\centering\includegraphics[width=8.2cm,clip]{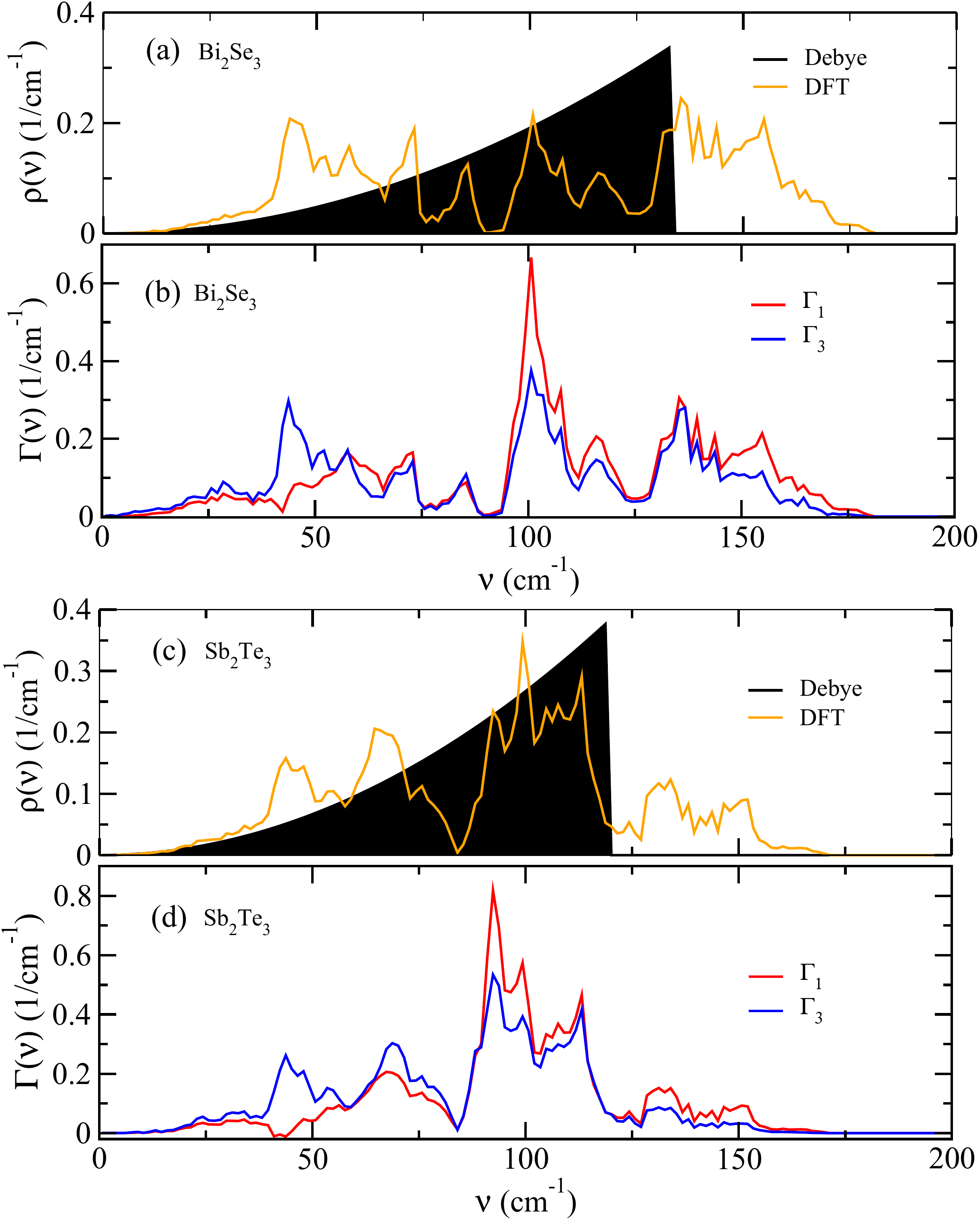}
\caption{
(a) The phonon density of states (from DFT and Debye approximation), $\rho(\nu)$. (b) The
phonon density of states weighted by \GP{}, $\Gamma_i(\nu)$,  due to an $xy$ biaxial strain and a $z$ uniaxial
strain for the \bise{}.  The corresponding results are shown in (c) and
(d) for \sbte{}.
}
\label{fig:vdos}
\end{figure}

For the above discussion, the temperature dependence of the TECs hinges
on the integrated quantities $I_i(T)$, which could be calculated from
a direct summation over BZ or through an integration over frequency
$\nu$ of the product of heat capacity $c(\nu,T)$ and the density of
phonon states weighted by \GP{}, $\Gamma_i(\nu)$. We find the second
approach is more illuminating. The density of phonon states for \bise{}
in \Fig~\ref{fig:vdos}(a) shows there is a phonon gap of $89 $ to
$94$~\invcm{} which is due mainly to a large mass difference between Bi
(atomic mass of 208.98) and Se (atomic mass of 78.97).  Such phonon
gap is not seen for \sbte{} [shown in \ref{fig:vdos}(c)] since Sb
(atomic mass of 121.76) and Te (atomic mass of 127.60) are consecutive
elements in the periodic table.  $\Gamma_i(\nu)$ for \bise{} are shown in
\Fig~\ref{fig:vdos}(b) for both $xy$ biaxial and $z$ uniaxial strains,
where large \GPs{} are associated with frequencies of about 100~\invcm.
For \sbte{}, large \GPs{} are associated with frequencies of about
90~\invcm.  $\Gamma_i(\nu)$ shown in \Fig~\ref{fig:vdos}(b) and (d) also
indicate that effect of negative \GPs{} are negligible for all frequencies
except for the $xy$ biaxial strain of \sbte{} at about 38~\invcm{}.

\begin{figure}[htbp]
\includegraphics[clip,width=8.0cm]{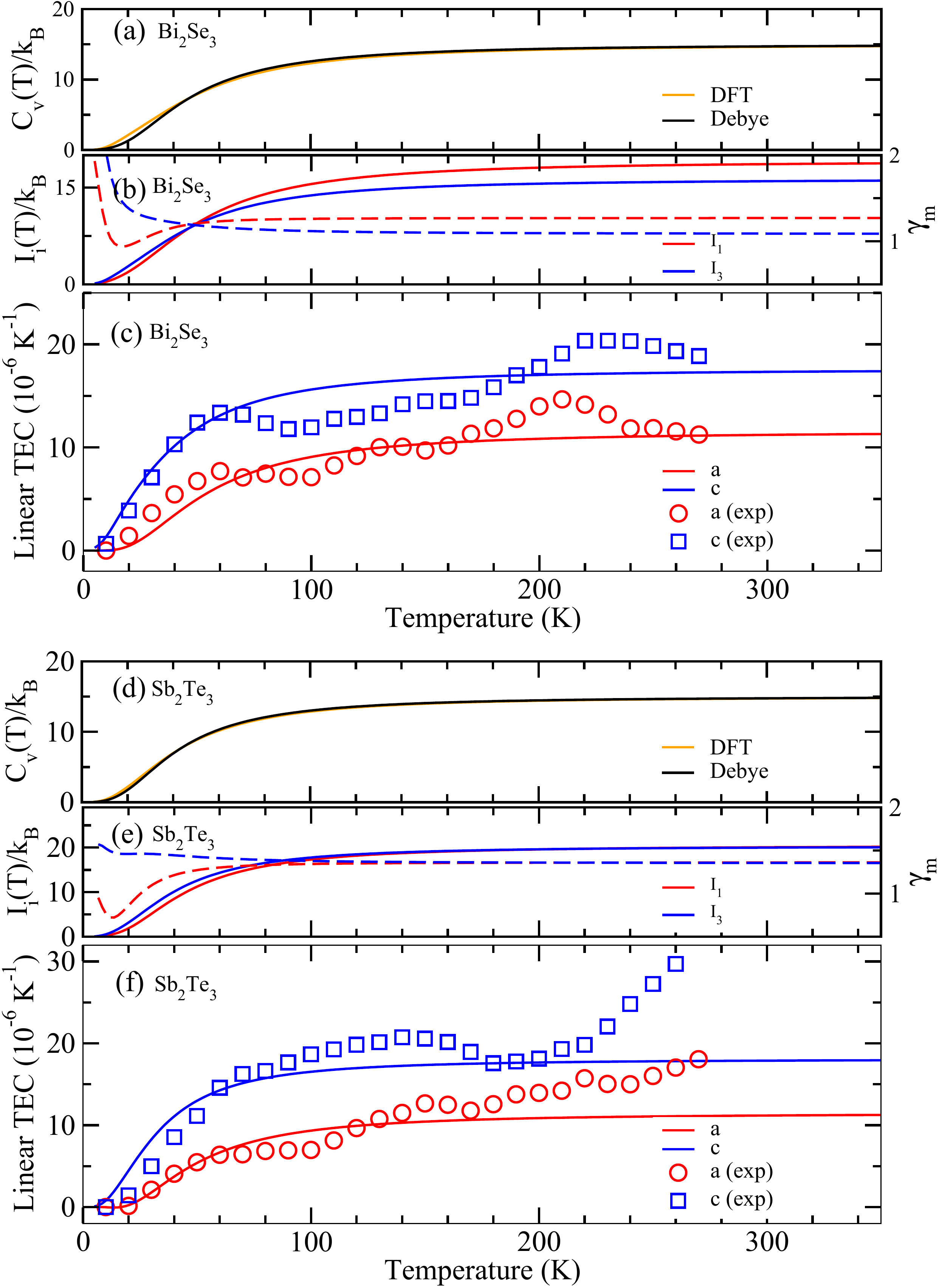}
\caption{
The temperature dependence of (a) $C_v$, (b) $I_i$ (solid lines) and
$\gamma_{m,i}$ (dashed lines), and (c) the linear TECs of \bise{}.
The respective data shown in (d), (e), and (f) are for \sbte{}.
}
\label{fig:ltec}
\end{figure}

Since the temperature dependence of TECs is intricately related to that of
the heat capacity at constant volume, which is typically characterized by
the Debye temperature, here we suggest a simple approach to extract the
effective Debye temperature.  From phonon calculations based on density-functional
perturbation theory (DFPT), we could obtain very accurate phonon
density of states and hence heat capacity as a function of temperature
[see \Fig~\ref{fig:ltec}(a) and (d)].  We propose to fit the DFT heat
capacity data with that obtained from a Debye model approximation by
minimizing the absolute error as a function of a cutoff frequency $\nu_c$,
\begin{equation}
d(\nu_c) = \frac{1}{(3Nk_B)^2} \int_{0}^{\infty} dT [ C^{D}_v(\nu_c,T) - C_v^{DFT}(T) ]^2 
\end{equation} 
where the integrand is the square of the difference between of the
heat capacities from DFT and from the Debye model.  $N =5$ is the
number of atoms in the primitive cell in current systems. According
to this scheme, the Debye temperature will be naturally defined as
$\theta_D = h\nu_c/k_B$. The heat capacity evaluated according to
the Debye model is $C^{D}_v(\nu_c,T) = \int_0^{\nu_c} \rho_D(\nu)
c(\nu,T) d\nu $ and the density of phonon states under the Debye
approximation is $\rho_D(\nu) = A \nu^2 $ for $ 0 \le \nu \le \nu_c$
and zero otherwise ($A = 9N/\nu_c^3$).  We note that a similar scheme
for finding the Debye temperature as a function of {\em temperature}
has been proposed in Ref.~[\onlinecite{Tohei06v73}].  The best
cutoff frequencies are 133 and 119~\invcm{} for \bise{} and \sbte{},
respectively. This translates to $\theta_D$ of 191~K and
172~K, respectively. These values are in the correct order and agree
well with the literature\cite{Madelung2004-book} values of 182 and
160~K, respectively.  It is interesting to see that even though the
phonon densities of states from the Debye approximation and DFT differ
significantly [see \Fig~\ref{fig:vdos}(a) and (c)], the heat capacities
between DFT and Debye approximation agree remarkably well with each other
[see \Fig~\ref{fig:ltec}(a) and (d)], which demonstrates the robustness
of Debye model to describe the heat capacity.

With the $\Gamma_i(\nu)$ data, we calculate the integrated quantities
$I_i(T)$ as shown in \Fig~\ref{fig:ltec}(b) and (e) (solid lines), for
\bise{} and \sbte{}, respectively. These are positive functions, which
eliminate the occurrence of negative linear TECs. For \sbte{}, $I_i(T)
$ for large-$T$ limit coincides fortuitously for $xy$ biaxial and $z$
uniaxial strains.  We also show the values of the macroscopic \GPs{},
$\gamma_{m,i}(T)$ in \Fig~\ref{fig:ltec}(b) and (e) (dashed lines).
\bise{} has a large-$T$ limit of $\gamma_{m}$ of 1.27 and 1.09 for
the $xy$ biaxial and $z$ uniaxial strains.  For \sbte{}, the large-$T$
limit of $\gamma_{m}$ is $1.36$ for both $xy$ biaxial and $z$ uniaxial
strains, which is in good agreement with a reported\cite{Stoffel15v27}
result of $1.40$. Therefore it is concluded that \sbte{} has a higher
phonon anharmonicity than \bise{} based on the macroscopic \GPs{}.

The linear TECs for \bise{} in the $a$ and $c$ directions are shown in
\Fig~\ref{fig:ltec}(c). We observe very good agreement between theory and experiment for $\alpha_c$ up to $60$~K, beyond
which the experiment data shows a dip between $60$ and $100$~K and
raises slowly after $100$~K.  The theoretical values for $\alpha_a$
are underestimated below $60$~K but a good agreement with experiment
is observed between $60$ and $180$~K.  The linear TECs for \sbte{}
in \Fig~\ref{fig:ltec}(f) show a reasonable agreement between theory
and experiment for both $\alpha_a$ and $\alpha_c$ for temperature
below $80$~K. Finally we note that the theoretical large-$T$ limit
of $\alpha_a$ for \bise{} and \sbte{} are the fortuitously the same
($11.3 \times 10^{-6}$~K$^{-1}$). The large-$T$ limit of $\alpha_c$
for \bise{} ($17.4 \times 10^{-6}$~K$^{-1}$) and \sbte{} ($17.3 \times
10^{-6}$~K$^{-1}$) are also very similar.  Since the elastic constants
for both materials are rather similar, from Eq.~1 we reason that the
slightly larger anharmonicity found in \sbte{} is somewhat compensated
by its slightly larger primitive cell volume, which results in very
similar temperature dependence of linear TECs for \bise{} and \sbte{}.
Finally, we note that the complicated temperature dependence of linear
TECs in experiments was argued to be attributed to higher-order
anharmonic effects and the breaking of the van der Waals bonds between
two Se-Se (or Te-Te) layers\cite{Chen11v99} at elevated temperatures.
We expect the use of quasi-harmonic approximation (QHA) may improve
the prediction of the \Gru{} formalism at higher temperatures,
however, we do not have enough computational resources for a full QHA treatment for both
crystals.

\section{Summary}
In summary, we have performed density-functional theory (DFT) calculations
to study the phonon anharmonicity of two trigonal systems \bise{} and
\sbte{}. Building upon previous computational approaches, we devised an efficient \Gru{} approach in calculating the linear
thermal expansion coefficients (TECs). The symmetry of the crystals are
fully utilized to reduce the comparatively expensive phonon calculations
(compared to standard DFT total-energy calculations) to a minimal set.
Even though the main aim of the paper is to study the linear TECs of the
systems, many intermediate quantities such as density of phonon states,
heat capacity, Debye temperature, mode-dependent \GP{}, density of
\GPs{}, density of phonon states weighted by \GP{}, and macroscopic \GP{}
have been carefully analyzed to shed light on the temperature dependence
of linear TECs of \bise{} and \sbte{}. Reasonably good agreement between
theory and experiment for linear TECs has been demonstrated. With the demonstrated accuracy and efficiency of the method,
we are confident that a wide applicability of our approach to
other thermoelectrics or even other classes of low-symmetry materials. 
We hope our results will encourage the inclusion of our method
in accelerated materials search packages.

The raw/processed data required to reproduce these findings cannot be shared at this time as the data also forms part of an ongoing study.

\section{Acknowledgments}
We gratefully thank the National Supercomputing Center (NSCC),
Singapore  and A*STAR Computational Resource Center (ACRC), Singapore
for computing resources.

\bibliographystyle{plain}
\bibliography{master-references}
\bibliographystyle{apsrev4-1}
\end{document}